\documentclass[journal]{IEEEtran}
\usepackage{amsthm}
\usepackage{amsfonts}
\usepackage[top=0.8in, bottom=1in, left=0.7in, right=0.7in]{geometry}
\usepackage{amssymb}
\ifCLASSINFOpdf
  \usepackage[pdftex]{graphicx}
  \usepackage{graphicx}
  \graphicspath{{../pdf/}{../jpeg/}{../png/}}
  \DeclareGraphicsExtensions{.pdf,.jpeg,.png}
\else
  \usepackage[dvips]{graphicx}
  \graphicspath{{../eps/}}
  \DeclareGraphicsExtensions{.eps}
\fi
\ifCLASSOPTIONcompsoc
  \usepackage[caption=false,font=normalsize,labelfont=sf,textfont=sf]{subfig}
\else
\usepackage[caption=false,font=footnotesize]{subfig}
\fi
\usepackage{multirow}
\usepackage{balance}
\usepackage{multicol}
\usepackage{epstopdf}
\usepackage{verbatim} 

\usepackage[english]{babel}
\usepackage{float}
\usepackage{xcolor}
\usepackage[linesnumbered,ruled,vlined]{algorithm2e}

\usepackage[hidelinks]{hyperref}
\usepackage{cite}
\ifCLASSINFOpdf
   \usepackage[pdftex]{graphicx}
   \else
\usepackage[dvips]{graphicx}
\fi
\usepackage[cmex10]{amsmath}
\usepackage[T1]{fontenc}
\newcommand*{\affmark}[1][*]{\textsuperscript{\dag}}
\begin{document}

\title{Ring Bayes Near-Field Channel Learning for mmWave Hybrid MIMO Systems Employing Uniform Circular Array}
\author{
\normalsize{Abhisha~Garg$^{*}$, Priya~Gupta$^{\dagger}$, Suraj~Srivastava$^{\dagger}$, Aditya~K.~Jagannatham$^{*}$,\\ Department of Electrical Engineering, Indian Institute of Technology Kanpur, India$^*$ \\ Department of Electrical Engineering, Indian Institute of Technology Jodhpur, India$^\dagger$\\(e-mail: abhisha20@iitk.ac.in$^*$; priyagupta@iitj.ac.in$^\dagger$; surajsri@iitj.ac.in$^\dagger$; adityaj@iitk.ac.in$^*$)}}
\maketitle
\thispagestyle{empty} 
\pagestyle{empty} 
\begin{abstract}
This work conceives a \textit{Ring-Bayes channel learning framework} that unifies Bayesian learning with near-field channel estimation in millimeter-wave (mmWave) hybrid MIMO systems. We develop a near-field channel model with a uniform circular array (UCA) at the base station (BS) and a uniform linear array (ULA) at the user equipment (UE), explicitly accounting for spherical wavefront propagation. Exploiting the inherent sparsity of mmWave channels in the joint angular–distance domain, we construct a \textit{structured sparse representation} and employ frame-wise processing to enhance the reliability of multi-antenna UE estimation. Furthermore, we design a concentric-ring codebook tailored to the UCA geometry, which efficiently captures near-field channel features across both angular and distance domains. Leveraging this structure, the proposed Ring-Bayes framework achieves highly accurate recovery of UCA near-field channels. Extensive simulations demonstrate significant performance gains over existing approaches, establishing Ring-Bayes as a scalable and powerful solution for next-generation mmWave communications.
\end{abstract}
\section{Introduction}\label{int_}
mmWave communication operating over the vast frequency band $30-300 \: \mathrm{GHz}$ \cite{heath2016overview}, has immense potential to meet the growing demand to support massive data traffic, offering low latency and unparalleled reliability. Massive multiple-input multiple-output (mMIMO) technology, which integrates a large number of antennas has been pivotal in enhancing spectrum efficiency by orders of magnitude. Although uniform linear arrays (ULAs) have been extensively considered in mmWave band, their deployment in practical mMIMO systems is inefficient, as they cannot accommodate a large number of antenna elements within the constrained device space, thereby limiting beamforming capabilities. In contrast, uniform circular arrays (UCAs) enable denser antenna placement \textit{within the same array area}, leading to higher transmission gain and enhanced data rates \cite{zhang20165g}. Furthermore, for large array aperture, the classical planar-wave assumption no longer holds, and the angular-domain sparsity alone fails to accurately recover the channels \cite{wei2021channel}. This limitation necessitates a shift towards \textit{near-field models}, where \textit{spherical-wavefront} are considered to more accurately characterize wave propagation. Cui and Dai \cite{cui2022channel}, in their seminal work proposed a near-field channel estimation framework  by exploiting polar-domain sparsity, enabling precise reconstruction of channel parameters in mMIMO systems. Zhang \textit{et al.} \cite{zhang2023near}, in their groundbreaking work, proposed a distance-parameterized angular-domain sparse representation model for near-field channel estimation, where sparse recovery is performed using orthogonal matching pursuit (OMP). Wu \textit{et al.} \cite{wu2024near}, proposed near-field channel estimation framework for dual-band extreme large (XL)-MIMO systems, where some side information is exploited to assist compressed sensing for improved recovery accuracy. However, none of the above mentioned works have considered UCA, which offer higher array gain and directivity compared to ULA, and demonstrate greater robustness to gain fluctuations and vibrations than Uniform Planar Arrays (UPA) \cite{zhang20165g}.

In \cite{zhang2019near}, Zhang and Fan introduced beamforming algorithm for UCAs, employing a successive cancellation scheme based on the phase mode excitation principle. Yang \textit{et al.} in \cite{yang2025near}, proposed a spherical-domain transform codebook that jointly models distance, azimuth, and elevation angles, and performed channel estimation using simultaneous OMP. However, it only explored single antenna user. Furthermore, OMP and its variants suffer from convergence issues and structural errors, thereby preventing UCAs from fully exploiting their inherent array gain and spatial resolution capabilities. Therefore, the novel contributions of this work are detailed next.
\subsection{Contributions}
We develop a comprehensive near-field channel model for mmWave hybrid MIMO systems with base station (BS) antennas arranged in a UCA fashion, while user antennas in ULA arrangement. Building on this, we introduce a frame-wise channel estimation model, which enhances estimation stability and accuracy under near-field conditions. Leveraging the inherent distance and angular sparsity of mmWave channels in near-field, we propose a concentric-ring codebook composed of beamfocusing vectors that concentrate energy on strategically sampled points. Sampling intervals are optimized by exploiting zeros of \textit{Bessel functions} to minimize coherence, thereby significantly enhancing sparse representation and enabling precise channel characterization. We further introduce the Ring Bayes channel learning framework, which integrates the tailored, \textit{low-coherence spherical-domain codebook with Bayesian principles} for improved channel estimation. We further demonstrate that this improved channel estimation also leads to enhanced bit-error-rate (BER) performance.
\subsection{Notation} \label{notation}
Upper-case letters $\mathbf{A}$ represent matrices, while lower-case letters $\mathbf{a}$ represent vectors. The quantity $\lfloor \cdot \rfloor$ represents the floor operation, while $|\cdot|$ represents the cardinality of a set. The identity $\mathrm{vec}(\mathbf{ABC}) = (\mathbf{C}^T \otimes \mathbf{A})\mathrm{vec}(\mathbf{B})$ is used to vectorize a matrix, where $\mathrm{vec}(\cdot)$ represents vectorization operation, while $\otimes$ represents the Kronecker product. The quantity $\lVert \cdot \rVert_{\mathcal{F}}$ represents the Frobenius norm, $\mathbb{E}\{\cdot\}$ represents the expectation operator, and $\mathcal{CN}(\boldsymbol{\mu}, \mathbf{\Sigma})$ represents a complex Gaussian distribution with mean $\boldsymbol{\mu}$ and covariance matrix $\mathbf{\Sigma}$.
\section{ System model}
\begin{figure}
\centering
\includegraphics[scale=0.27]{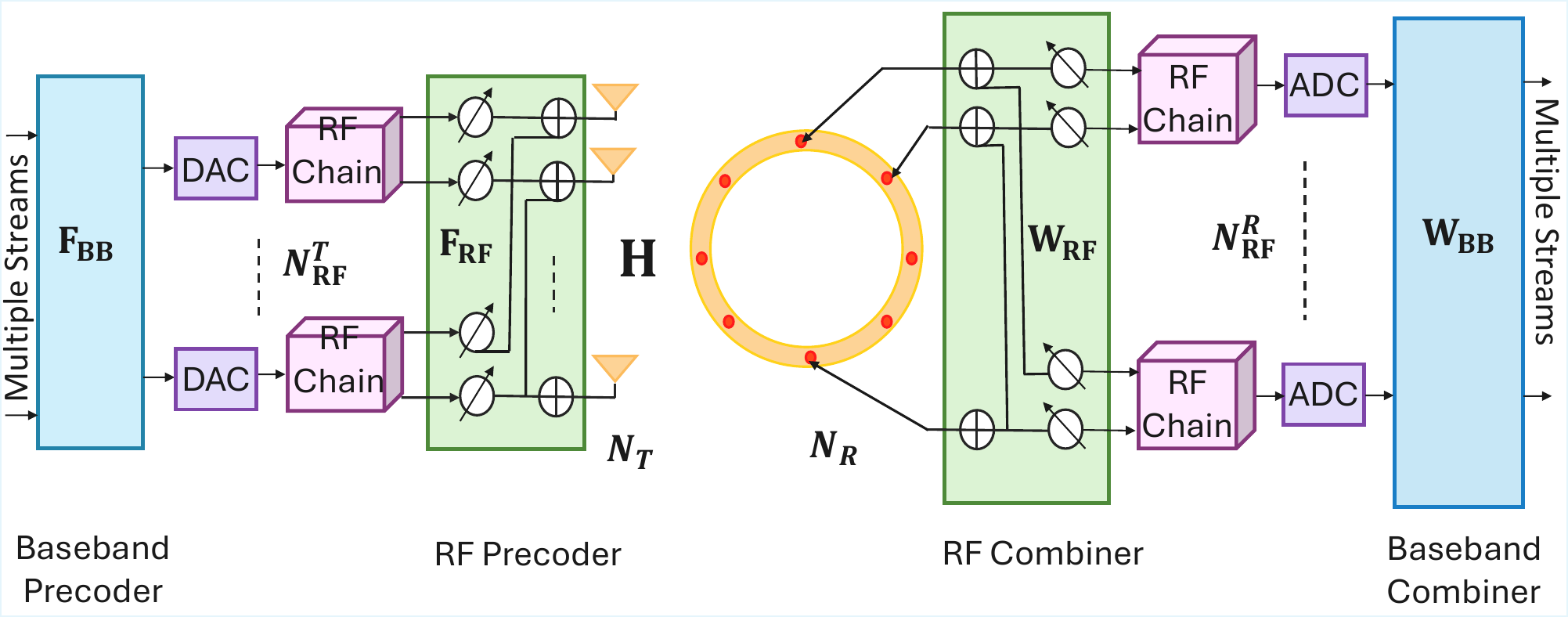}
\caption{Block diagram of mmWave hybrid MIMO system}
\label{system}
\end{figure}
Consider a mmWave hybrid mMIMO system, where the base station (BS) is equipped with a UCA and the user is equipped with a ULA, as illustrated in Fig. \ref{system}. The BS has $N_R$ antennas with $N_{\mathrm{RF}}^R$ RF chains while the user has $N_T$ antennas with $N_{\mathrm{RF}}^T$ RF chains and $N_s$ data streams to transmit. The hybrid architecture includes baseband and RF modules at both BS and user, where the baseband and RF precoders at the user are given as $\mathbf{F}_{\mathrm{BB}} \in \mathbb{C}^{N_{\mathrm{RF}}^T \times N_s}$ and $\mathbf{F}_{\mathrm{RF}} \in \mathbb{C}^{N_T \times N_{\mathrm{RF}}^T}$, respectively, while the baseband and RF combiners at BS are given as $\mathbf{W}_{\mathrm{BB}} \in \mathbb{C}^{N_{\mathrm{RF}}^R \times N_s}$ and $\mathbf{W}_{\mathrm{RF}} \in \mathbb{C}^{N_R \times N_{\mathrm{RF}}^R}$. Moreover, the RF precoder and combiner follow unit modulus constraint given as $|\mathbf{F}_{\mathrm{RF}}(\imath,\jmath)| = \frac{1}{\sqrt{N_T}}$ and $|\mathbf{W}_{\mathrm{RF}}(\imath,\jmath)| = \frac{1}{\sqrt{N_R}}$ respectively. Let $\mathbf{x} \in \mathbb{C}^{N_s \times 1}$ represent the transmit symbol vector following the constraint $\mathbb{E}\{{\mathbf{x}\mathbf{x}^H}\} = \mathbf{I}_{N_s}$. In OFDM system, at the transmitter, the symbols associated with each RF chain undergo K-point inverse fast Fourier transform (IFFT) processing and then a cyclic prefix (CP) of length $L$ is appended. At the receiver, the incoming signal  at each RF chain is processed by removing the CP and transformed back to the frequency domain using $K$ point FFT. Therefore, the received signal $\mathbf{y} \in \mathbb{C}^{N_s \times 1}$ after RF and baseband combining can be given as
\begin{align}
{\mathbf{y}} = \mathbf{W}_{\mathrm{BB}}^H\mathbf{W}_{\mathrm{RF}}^H\mathbf{H}\mathbf{F}_{\mathrm{RF}}\mathbf{F}_{\mathrm{BB}}\mathbf{x} + \mathbf{W}_{\mathrm{BB}}^H\mathbf{W}_{\mathrm{RF}}^H\check{\mathbf{n}}, \label{final_freq_output}
\end{align}
where $\check{\mathbf{n}} \in \mathbb{C}^{N_R \times 1}$ represents the additive white Gaussian noise which follows the distribution $\mathcal{CN}(0,\sigma^2)$ and $\mathbf{H} \in \mathbb{C}^{N_R \times N_T}$ represents the mmWave MIMO channel, formulation of which is described in Section-\ref{channel_model}. Next we develop a pilot based channel estimation model.

Note that, during the training process, the baseband unit cannot be optimized due to the lack of CSI. Hence during channel estimation, in the $m$-th pilot frame, we directly transmit a pilot vector $\mathbf{x}_m \in \mathbb{C}^{N_{\mathrm{RF}}^T \times 1}$ using RF precoder $\mathbf{F}_{\mathrm{RF},m}$ and combine the received output using the combiner $\mathbf{W}_{\mathrm{RF},m}$. After applying the $\mathrm{vec}(\cdot)$ operator property as described in Section-\ref{notation}, the combined pilot output $\mathbf{y}_m \in \mathbb{C}^{N_{\mathrm{RF}}^R \times 1}$ at $m$-th time instant can be re-written as
\begin{align}
\mathbf{y}_m = \underbrace{(\mathbf{x}_m^T\mathbf{F}_{\mathrm{RF},m}^T\otimes \mathbf{W}_{\mathrm{RF},m}^H)}_{\tilde{\mathbf{\Omega}}_m \in \mathbb{C}^{N_{\mathrm{RF}}^R\times N_R N_T}}\mathbf{h} + {\mathbf{n}}_m, \label{recieved_eq}
\end{align}
where $\mathbf{h} = \mathrm{vec}(\mathbf{H}) \in \mathbb{C}^{N_RN_T\times 1}$ represents the vectorized mmWave MIMO channel and $\tilde{\mathbf{\Omega}}_m$ represents the sensing matrix. Let $\mathbf{n}_m = \mathbf{W}_{\mathrm{RF},m}^H\check{\mathbf{n}}_m\in \mathbb{C}^{N^R_{\mathrm{RF}}\times 1}$ represent the equivalent noise for notational simplicity. Furthermore, the noise covariance $\mathbf{E}\{\mathbf{n}_m\mathbf{n}_m^H\} = \sigma^2\mathbf{W}_{\mathrm{RF},m}^H\mathbf{W}_{\mathrm{RF},m} = \mathbf{S}_m$. The equivalent channel estimation model after concatenating $M$ pilot blocks yields
\begin{align}
\underbrace{\begin{bmatrix}\mathbf{y}_1\\ \vdots\\ \mathbf{y}_M\end{bmatrix}}_{\mathbf{y}_p} = \underbrace{ \begin{bmatrix} \tilde{\mathbf{\Omega}}_1\\ \vdots\\ \tilde{\mathbf{\Omega}}_M\end{bmatrix}}_{\tilde{\mathbf{\Omega}}_p}\mathbf{h} + \underbrace{\begin{bmatrix} \mathbf{n}_1\\ \vdots\\ \mathbf{n}_M\end{bmatrix}}_{\mathbf{n}_p}, \label{con_frame}
\end{align}
where $\mathbf{y}_p \in \mathbb{C}^{MN_{\mathrm{RF}}^R \times 1}$ represents the stacked pilot output, $\tilde{\mathbf{\Omega}}_p \in \mathbb{C}^{MN_{\mathrm{RF}}^R \times N_R N_T}$ denotes the stacked pilot sensing matrix and $\mathbf{n}_p \in \mathbb{C}^{MN_{\mathrm{RF}}^R \times 1}$ represents the stacked noise. Moreover, the $\mathbb{E}\{\mathbf{n}_p\mathbf{n}_p^H\} = \mathrm{blkdiag}(\mathbf{S}_1, \mathbf{S}_2, \cdots, \mathbf{S}_M) = \mathbf{S} \in \mathbb{C}^{MN_{\mathrm{RF}}^R \times MN_{\mathrm{RF}}^R}$. The next subsection will discuss the near-field MIMO channel model.
\begin{figure}
\centering
{\includegraphics[scale=0.25]{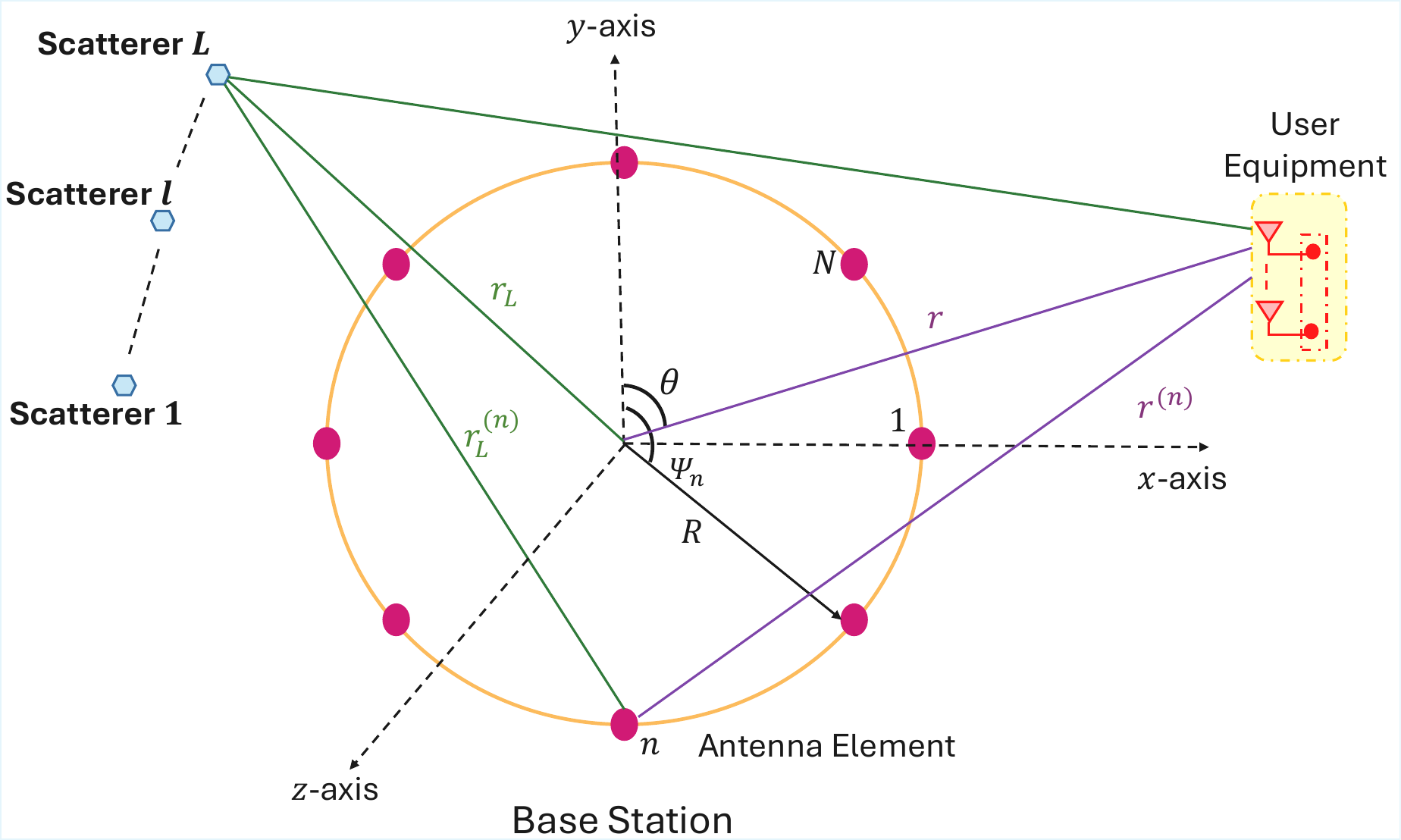}}
\caption{Geometric relation between UCA and user in near-field with multipath propagation environment}
\label{UCA}
\end{figure}
\subsection{Near-Field Channel Model} \label{channel_model}
A far-field beam steering vector $\mathbf{a}(\theta) \in \mathbb{C}^{N \times 1}$ for UCA of BS is given as \cite{wu2023enabling}
\begin{align}
    \check{\mathbf{a}}(\theta) =\frac{1}{\sqrt{N}}[e^{-j\frac{2 \pi}{\lambda}R\cos(\theta-\psi_1)}, \cdots, e^{-j\frac{2 \pi}{\lambda}R\cos(\theta-\psi_N)}]^T, \label{far_array}
\end{align}
where $R$ represents the radius of the circle on which the antennas are uniformly distributed, as shown in Fig. \ref{UCA}, and $\psi_n$ represents the antenna location defined as $\psi_n = \frac{2 \pi n}{N}, \: \forall \: n= 1,2 ,\cdots N$. However, as detailed in Section-\ref{int_}, for a large antenna array at BS, the far-field assumption is no longer valid, and a \textit{spherical-wave} propagation model has to be considered. Therefore, the array steering vector for near-field propagation can be rewritten as \cite{zhang2019near}
\begin{align}
\mathbf{a}(r,\theta)
= \frac{1}{\sqrt{N}}\big[e^{-j \frac{2\pi}{\lambda}(r^{(1)}-r)},
\cdots, e^{-j \frac{2\pi}{\lambda}(r^{(N)}-r)}\big]^T,
\end{align}
As depicted in Fig. \ref{UCA}, 
$r^{(n)}$ indicates the propagation distance from the user to the
$n$th antenna of the UCA. This distance can be mathematically defined as
\begin{equation}
    \begin{aligned}
         r^{(n)} &= \sqrt{r^{2} + R^{2} - 2rR \cos(\theta-\psi_{n})}\\ & \approx r - R \cos(\theta - \psi_{n}) + \frac{R^{2}}{2 r}[1 - \cos^{2}(\theta - \psi_{n})].
    \end{aligned}
\end{equation}
Note that the approximation in the above equation is derived from the \textit{second-order Taylor series} expansion $\sqrt{1+x} = 1+\frac{x}{2}-\frac{x^2}{8}+\mathcal{O}(x^3)$ assuming $r$ is large compared to the other terms \cite{wu2023enabling}. Therefore, the mmWave MIMO channel can be modeled as
\begin{equation}
\mathbf{H}= \sqrt{\frac{N_T N_R}L} \sum_{l=1}^{L}\alpha_l\,
\mathbf{a}_R(r_l,\theta_l)\tilde{\mathbf{a}}_T^H(\phi_l).
\end{equation}
Here $L$ represents the number of scatterers and $\tilde{\mathbf{a}}(\phi) \in \mathbb{C}^{N_T \times 1}$ represents the beam steering vector of ULA at user which is given by
\begin{equation}
\tilde{\mathbf{a}}(\phi)= \frac{1}{\sqrt{N_T}}\big[1,
e^{-j \frac{2\pi}{\lambda} d\cos\phi},\;
\cdots,e^{-j \frac{2\pi}{\lambda} (N_T - 1) d \cos\phi}\big]^T,
\end{equation}
where $d$ represents the antenna spacing in ULA. Note that, since $N_T \ll N_R$, we assume far-field propagation from scatterer to the user. The subsequent section describes the development of a concentric-ring codebook tailored for uniform circular arrays, which will be subsequently utilized to formulate a sparse channel estimation model.
\section{Sparse channel estimation model}
As discussed in Section-\ref{int_}, near-field propagation exhibits joint angle-range dependence, affecting the spatial propagation characteristics. Therefore, the beamforming gain be given as \eqref{beam_gain}.
\begin{figure*}
    \begin{equation}
    \begin{aligned}
    g(r_1,\theta_1,r_2,\theta_2) &= \big|\mathbf{a}^H(r_1,\theta_1)\mathbf{a}(r_2,\theta_2)\big| \\ & = \frac{1}{N} \Big|\sum_{n=1}^N e^{j \frac{2 \pi}{\lambda}\Big(\sqrt{r_1^2+R^2-2r_1R \cos(\theta_1-\psi_n)-r_1}\,\Big)} e^{-j\frac{2 \pi}{\lambda}\Big(\sqrt{r_1^2+R^2-2r_1R \cos(\theta_1-\psi_n)-r_1}\, \Big)}  \Big| \\ &\approx \frac{1}{N}\bigg|\sum_{n=1}^N e^{j\frac{2 \pi}{\lambda}R(\cos(\theta_1-\psi_n)-\cos(\theta_2-\psi_n))}\times  e^{-j\frac{2 \pi}{\lambda}R^2\Big(\frac{1-\cos^2(\theta_1-\psi_n)}{2 r_1}-\frac{1-\cos^2(\theta_2-\psi_n)}{2 r_2}\Big)}\Big|. \label{beam_gain}
\end{aligned}
\end{equation}
\hrulefill
\end{figure*}
Moreover, to effectively calculate the above quantity, we fix one parameter and vary the other as discussed below.
\begin{figure}
\centering
\subfloat[]{\includegraphics[scale=0.3]{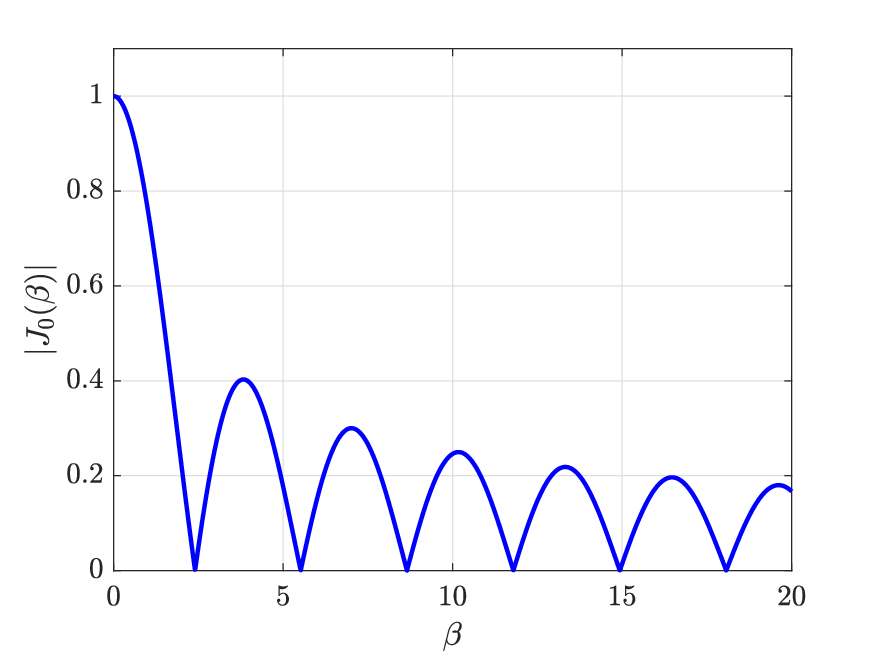}}
 \hfil
\hspace{-10pt} \subfloat[]{\includegraphics[scale=0.21]{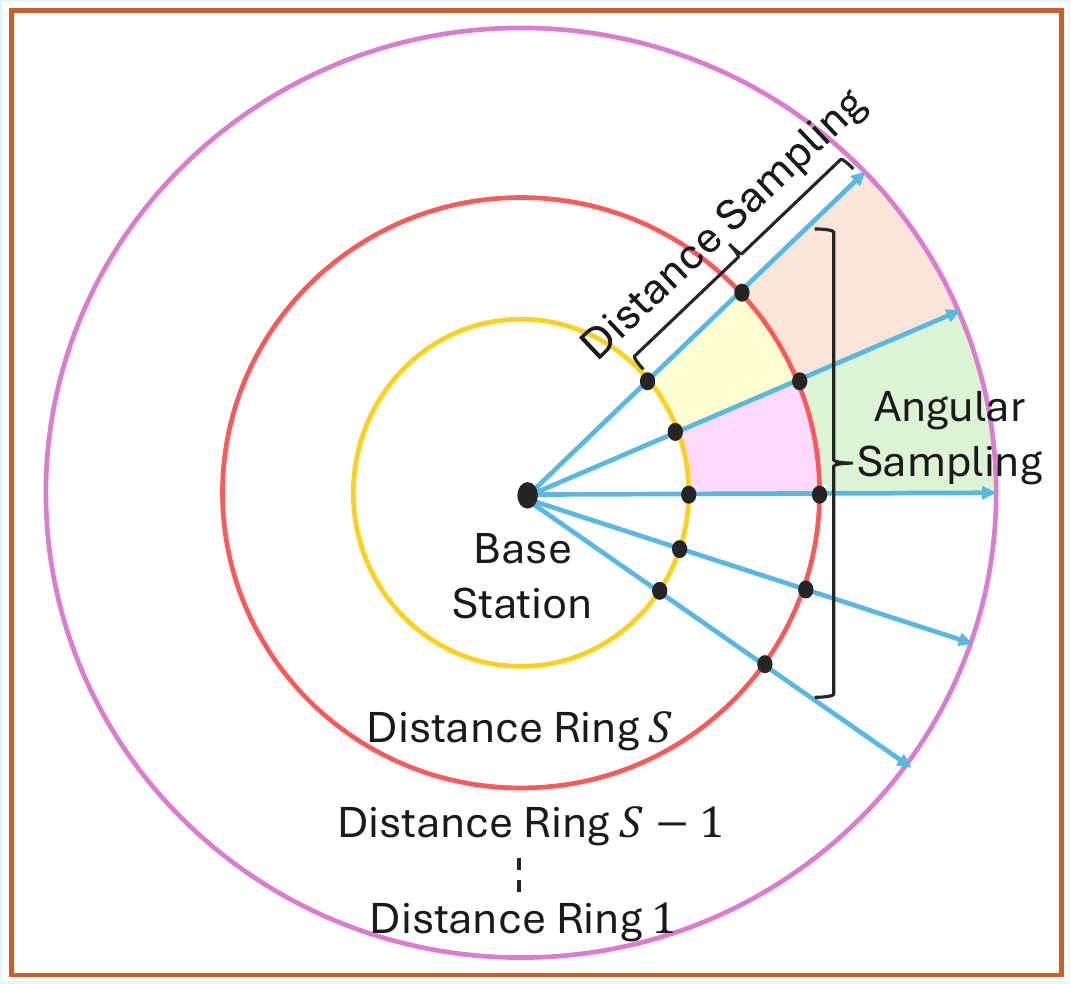}}
\caption{$ \left(a\right) $ Illustration of the magnitude of the zeroth-order Bessel function $ \left(b\right) $ Concentric-ring codebook illustration for UCA}
\label{besl}
\end{figure}
\subsection{Angular Domain Sampling} \label{an_sample}
In order to achieve the beamforming gains using near-field beamforming vectors, we assume $r_1 = r_2 = r$ in Eq. \eqref{beam_gain}, which on simplification yields
\begin{align}
    g(r,\theta_1,r,\theta_2) \approx \frac{1}{N} \Big|\sum_{n=1}^N e^{j\beta \sin \big(\psi_n-\frac{\theta_1+\theta_2}{2}\big)}\Big|,
\end{align}
where $
\beta = \frac{4\pi R}{\lambda} \sin(\frac{\theta_{2} - \theta_{1}}{2})$. Moreover, using the \textit{Jacobi-Anger} expansion \cite{bowman2012introduction}, which states $e^{j \beta \cos\gamma} = \sum_{m=-\infty}^\infty j^m J_m(\beta) e^{jm\gamma},$ where $J_m(.)$ denotes the $m$-th order Bessel function; it can be observed that the beamforming gain can be expressed as a summation of Bessel functions of different orders. However, due to the orthogonality and asymptotic properties of Bessel functions, contributions from higher-order terms become negligible for large antenna arrays. Consequently, the beamforming gain is predominantly governed by the \textit{zeroth-order Bessel function} $J_0(\beta),$ which captures the main spatial response of the array.

From Fig. \ref{besl}(a), one can further observe that the beamforming gain in the angular domain exhibits oscillatory behavior. For instance, when $\beta \in [0,\, 2.4048]$ the beamforming gain decreases monotonically, which is desirable. However, for $\beta \in [ 2.4048 ,\,4 ]$, the beamforming gain increases again. Therefore, by setting the angular spacing between two adjacent codewords, such that $\beta \approx 2.4048$, corresponding to the first zero of $|J_{0}(\beta)|$, this undesired gain increase can be mitigated. Furthermore, a suitable value corresponding to angular sampling can be considered as $
g(r, \theta_{1}, r, \theta_{2}) \approx \left| J_{0}(\beta) \right| \leq \Delta$, where $\Delta$ represents the threshold. From Fig. \ref{besl}(a), one can further observe that $|J_0(.)|$ is monotonically decreasing within its main lobe. Hence, the angular separation between adjacent samples can be determined such that their correlation equals $\Delta$, which is given by $\theta_{\Delta} = 
2 \sin^{-1}\big(\frac{\lambda J_{0}^{-1}(\Delta)}{4 \pi R}\big)$.
Therefore, the angular samples as described in Fig. \ref{besl}(b) are obtained as
\begin{equation}
\theta_{s_{1}} = s_{1}\theta_{\Delta}, \quad s_{1} = 0, 1, 2, \ldots, S_{1},
\end{equation}
where $
S_{1} = \big \lfloor \frac{2\pi}{\theta_{\Delta}} \big\rfloor - 1$, which ensures $\theta_{s_{1}} \in [0, 2\pi)$.
\subsection{Distance Domain Sampling}
To analyze the beamforming performance with respect to distance, we consider $\theta_1=\theta_2 =\theta$ in Eq. \eqref{beam_gain}, which allows the beamforming gain to be approximated as
\begin{equation}
g(r_{1}, \theta, r_{2}, \theta) \approx \frac{1}{N}\Big|\sum_{n=1}^N e^{j\zeta\cos(2 \theta-2 \psi_n)}\Big|,
\end{equation}
where $\zeta = \frac{2 \pi R^{2}}{\lambda} 
\big( \frac{1}{4r_{1}} - \frac{1}{4r_{2}} \big)$. Using the \textit{Jacobi-Anger} expansion of Bessel functions, as done in Section-\ref{an_sample}, the beamforming gain can be expressed as a summation of Bessel functions of different orders, where the \textit{zeroth-order Bessel function} contributes predominantly to the main spatial response and can be approximated as $J_0(\zeta)$. Therefore, in order to obtain accurate distance samples for threshold $\Delta$, the correlation of samples should follow the constraint $\big|\frac{1}{r_p}-\frac{1}{r_q} \big| \geq \frac{\lambda J_0^{-1}(\Delta)}{2 \pi R^2}$. Hence, the distance samples, as described in Fig. \ref{besl}(b), are obtained as
\begin{equation}
r_{s_{2}} = \frac{1}{s_{2}} \cdot \frac{ \pi R^{2}}{2 \lambda J_{0}^{-1}(\Delta)}, \quad s_{2} = 0, 1, \ldots, S_{2},
\end{equation}
where $S_{2} = \big \lfloor\frac{\pi R^{2}}{2 \lambda J_{0}^{-1}(\Delta)r_{\min}} \big\rfloor$ and $r_\mathrm{min}$ represents the minimum communication distance, i.e., Fresnel distance. Algorithm-\ref{alg:codebook} summarizes the procedure for concentric-ring based codebook generation.

At this juncture, it is important to note that the mmWave signal possesses a highly directional nature, and due to the presence of only a few dominant scatterers, the near-field mmWave MIMO channel exhibits sparsity in both angular and distance domains. We next discuss the sparse channel estimation model formulation.
\begin{algorithm}[t]
\caption{Design of a Near-Field Concentric-Ring Codebook $\mathbf{A}_R(r,\theta)$}
\label{alg:codebook}
\KwIn{$r_{\min}$, $\Delta$, $R$, $\lambda$}
\KwOut{Concentric-ring codebook $\mathbf{A}_R(r,\theta) \in \mathbb{C}^{N_R \times |\kappa_\theta||\kappa_r|}$}

\textbf{Initialization:} $s_{1} = 0, s_{2} = 0,$ $\theta_{\Delta} = 2 \sin^{-1}\big(\dfrac{\lambda {J}_{0}^{-1}(\Delta)}{4 \pi R}\big)$, 
   $r_{\Delta} = \dfrac{\pi R^{2}}{2 \lambda {J}_{0}^{-1}(\Delta)},$ $\kappa_{\theta} = \{0, 1, \ldots, \lfloor \frac{2\pi}{\theta_{\Delta}} \rfloor\}$,
   $\kappa_{r} = \{0, 1, \ldots, \lfloor \tfrac{r_{\Delta}}{r_{\min}} \rfloor\}$

\For{$s_{1} \in \kappa_{\theta}, \; s_{2} \in \kappa_{r}$}{
   $\theta_{s_{1}} = s_{1} \theta_{\Delta}$,
   $r_{s_{2}} = s_{2} r_{\Delta}$
}

$\mathbf{A}_R(r,\theta) = \{ \mathbf{a}(r_{s_{2}}, \theta_{s_{1}})\mid s_{1} \in \kappa_{\theta},s_{2} \in \kappa_{r}\}$

\end{algorithm}
Since we consider a ULA at the user under the far-field assumption; it, only needs angular-domain sampling. Accordingly, the user-side angular grid is formed by considering the directional cosines, which are uniformly spaced and given as
$\Phi_T = \big\{\phi_t:\cos(\phi_t)=\frac{2}{G_{T}}(t-1)-1,\; 1 \leq t \leq G_{T} \big\}$, where $G_T$ represents angular grid size. Moreover, user dictionary matrix $\tilde{\mathbf{A}}(\Phi_T) \in \mathbb{C}^{N_T \times G_T}$ can be given as
\begin{equation}
\tilde{\mathbf{A}}(\Phi_T) = \big[\tilde{\mathbf{a}}_T(\phi_1),\tilde{\mathbf{a}}_T(\phi_2), \cdots, \tilde{\mathbf{a}}_T(\phi_{G_{T}})\big].
\end{equation}
Thus, the mmWave hybrid MIMO channel can be well approximated as
\begin{align}
\mathbf{H} \approx \mathbf{A}_R(r,\theta) {\mathbf{H}}_b\tilde{\mathbf{A}}_T^H(\Phi_T), \label{freq_beam_channel}
\end{align}
where $\mathbf{H}_b \in \mathbb{C}^{\lvert \kappa_\theta \rvert \lvert \kappa_r \rvert \times \mathit{G}_T}$ represents an equivalent beamspace-domain channel, which is sparse in nature. By leveraging the $\mathrm{vec}(.)$ operator property, as discussed in Section-\ref{notation}, Eq. \eqref{freq_beam_channel} can be recast as
\begin{align}
\mathbf{h} = \textrm{vec}(\mathbf{H}) = \underbrace{[\tilde{\mathbf{A}}^{*}_T(\Phi_T) \otimes \mathbf{A}_{\mathit{R}}(r,\theta)]}_{\mathbf{\Lambda}}\mathbf{h}_b, \label{vec_channel_freq}
\end{align}
where $\mathbf{h}_{\mathit{b}} = \mathrm{vec}(\mathbf{H}_b) \in \mathbb{C}^{\lvert \kappa_\theta \rvert \lvert \kappa_r \rvert G_T \times 1}$ and $\mathbf{\Lambda} = [\tilde{\mathbf{A}}^{*}_T(\Phi_T) \otimes \mathbf{A}_R(r,\theta)] \in \mathbb{C}^{N_RN_T \times \lvert \kappa_\theta \rvert \lvert \kappa_r \rvert G_T}$ denotes a \textit{sparsifying-dictionary}. By substituting Eq. \eqref{vec_channel_freq} in \eqref{con_frame}, the received signal can be given as
\begin{align}
\mathbf{y}_{\mathit{p}} = \mathbf{\Omega}\mathbf{h}_b + \mathbf{n}_p, \label{vec_pilot_beam}
\end{align}
where $\mathbf{\Omega} = \tilde{\mathbf{\Omega}}_p\mathbf{\Lambda} \in \mathbb{C}^{MN_{\mathrm{RF}}^R \times G_RG_T}$ represents the equivalent sensing matrix. Due to the inherent sparsity of mmWave channels, they enable compressive sensing (CS) techniques to efficiently exploit sparse structure, allowing accurate channel estimation with substantially fewer pilot measurements compared to conventional methods. In this regard, the proposed ring Bayes algorithm leverages statistical priors and observed data for enhancing estimation accuracy. The next section develops the proposed Ring Bayes channel learning framework.
\section{Ring Bayes Channel Learning Framework}
Let a parameterized Gaussian prior $g(\mathbf{h}_b;\mathbf{\Gamma})$ is assigned to $\mathbf{h}_{b}$ as
\begin{align}
    g(\mathbf{h}_b; \mathbf{\Gamma}) = \prod_{i=1}^{\lvert \kappa_\theta \rvert \lvert \kappa_r \rvert G_T} (\pi \gamma_i)^{-1} \: \mathrm{exp}\Big(-\frac{|\mathbf{h}_b(i)|^2}{\gamma_i}\Big), \label{prior}
\end{align}
where $\mathbf{\Gamma} = \mathrm{diag}(\gamma_1,\cdots,\gamma_{\lvert \kappa_\theta \rvert \lvert \kappa_r \rvert G_T})$ represents the hyperparameter matrix and $\gamma_i$ represents the $i$-th hyperparameter. The log-likelihood of hyperparameter matrix can be formulated as
\begin{align}
    \log[g\,(\mathbf{y}_p;\mathbf{\Gamma})] = -MN_{\mathrm{RF}}^R\mathrm{log}(\pi)-\log[\mathrm{det}(\mathbf{R}_y)]-  \notag \\\mathbf{y}_p^H\mathbf{R}_y^{-1}\mathbf{y}_p,
\end{align}
where $\mathbf{R}_y = \mathbb{E}\{\mathbf{y}_p \mathbf{y}_p^H\} = \mathbf{S}+\mathbf{\Omega}\mathbf{\Gamma}\mathbf{\Omega}^H$. However, maximizing the objective above with respect to $\mathbf{\Gamma}$ cannot directly provide a closed-form solution. Therefore, we adopt the EM framework, as it ensures local convergence. Let $\hat{\gamma}_i^{(j-1)}$ represent the $i$th hyperparameter estimate in $(j-1)$th EM iteration and let $\hat{\mathbf{\Gamma}}^{(j-1)}=\mathrm{diag}\:(\gamma_1^{(j-1)},\cdots,\gamma_{\lvert \kappa_\theta \rvert \lvert \kappa_r \rvert G_T}^{(j-1)})$. During E-step, we maximize the log-likelihood of the complete dataset $\mathcal{L}(\mathbf{\Gamma}|\hat{\mathbf{\Gamma}}^{(j-1)})$ given by
\begin{equation}
    \begin{aligned}
    \mathcal{L}(\mathbf{\Gamma}|\hat{\mathbf{\Gamma}}^{(j-1)}) = & \mathbb{E}_{\mathbf{h}_b|\mathbf{y}_p;\hat{\mathbf{\Gamma}}^{(j-1)}}\{\log g[(\mathbf{y}_p,\mathbf{h}_b; \mathbf{\Gamma})]\} \\
     = &\mathbb{E}_{\mathbf{h}_b|\mathbf{y}_p;\hat{\mathbf{\Gamma}}^{(j-1)}}\{\log[g \,(\mathbf{y}_p|\mathbf{h}_b)]\} +  \\ &\mathbb{E}_{\mathbf{h}_b|\mathbf{y}_p;\hat{\mathbf{\Gamma}}^{(j-1)}}\{\log[g \,(\mathbf{h}_b;\mathbf{\Gamma})]\}.
\end{aligned}
\end{equation}
Upon simplification, the first term inside the $\mathbb{E}\{.\}$ operator is independent of hyperparameter matrix $\mathbf{\Gamma}$ and thus can be ignored in the subsequent M-step. Therefore, the equivalent optimization problem for M-step can be given as
\begin{align}
    \hat{\mathbf{\Gamma}}^{(j)} = \mathop{\mathrm{arg \: max}}_{\mathbf{\Gamma}} \: \mathbb{E}_{\mathbf{h}_b|\mathbf{y}_p;\hat{\mathbf{\Gamma}}^{(j-1)}} \{\mathrm{log}[g(\mathbf{h}_b;\mathbf{\Gamma})]\}.
\end{align}
Substituting Eq. \eqref{prior} in above Eq., differentiating it w.r.t $\gamma_i$ and setting the resultant to $0$, the individual hyperparameter estimate $\hat{\gamma}_i$ can be given by
\begin{align}
    \hat{\gamma}_i^{(j)} = \mathbb{E}_{\mathbf{h}_b|\mathbf{y}_p;\hat{\mathbf{\Gamma}}^{(j-1)}}\{|\mathbf{h}_b(i)|^2\}, \label{hype}
\end{align}
where the \textit{a posteriori} distribution of $\mathbf{h}_b$ can be expressed as $g(\mathbf{h}_b|\mathbf{y}_p;\hat{\mathbf{\Gamma}}^{(j-1)}) = \mathcal{CN}(\boldsymbol{\mu}^{(j)},\mathbf{\Sigma}^{(j)})$ and is given by
\begin{align}
    \boldsymbol{\mu}^{(j)} = \mathbf{\Sigma}^{(j)}\mathbf{\Omega}^H\mathbf{S}^{-1}\mathbf{y}_p, \: \: \mathbf{\Sigma}^{(j)} = \big[\mathbf{\Omega}^H\mathbf{S}^{-1}\mathbf{\Omega}+(\hat{\mathbf{\Gamma}}^{(j-1)})^{-1}\big]^{-1}. \label{mean_co}
\end{align}
The quantity $\boldsymbol{\mu}^{(j)} \in \mathbb{C}^{\lvert \kappa_\theta \rvert \lvert \kappa_r \rvert G_T \times 1}$ represents \textit{a posteriori} mean and $\mathbf{\Sigma}^{(j)} \in \mathbb{C}^{\lvert \kappa_\theta \rvert \lvert \kappa_r \rvert G_T \times \lvert \kappa_\theta \rvert \lvert \kappa_r \rvert G_T}$ represents \textit{a posteriori} covariance of the beamspace channel. After substituting Eq. \eqref{mean_co} in Eq. \eqref{hype}, the hyperparameter estimate can be simplified to
\begin{align}
    \hat{\gamma}_i^{(j)} = \mathbf{\Sigma}^{(j)}(i,i) + |\boldsymbol{\mu}^{(j)}(i)|^2. \label{hyper_para}
\end{align}
\begin{figure*}
	\centering
   \subfloat[]{\includegraphics[scale=0.42]{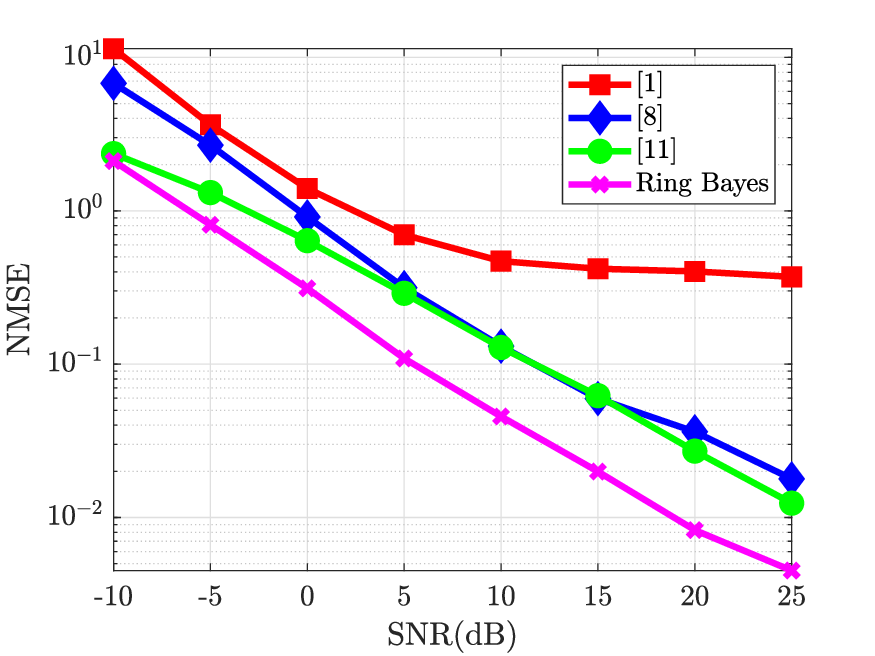}}
	\hfil
\hspace{-10pt}\subfloat[]{\includegraphics[scale=0.42]{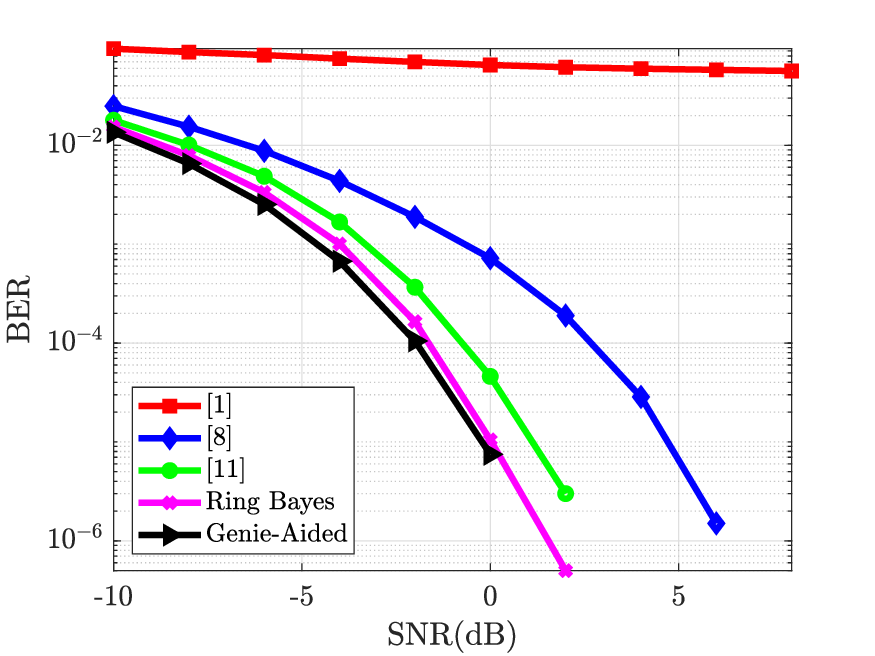}}
	\hfil
\hspace{-10pt}\subfloat[]{\includegraphics[scale=0.42]{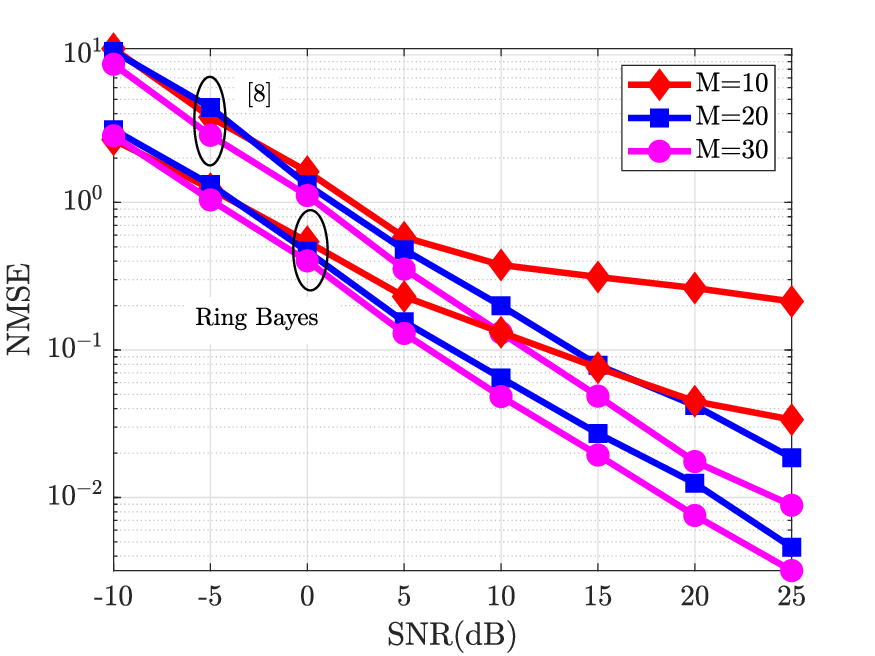}}
	\caption{$ \left(a\right) $ NMSE vs SNR comparison for the proposed and existing state-of-the-art approaches $ \left(b\right) $ BER vs SNR comparison for the proposed and existing state-of-the-art approaches $ \left(c\right) $ NMSE vs SNR comparison by varying pilot blocks $M$.}\label{NMSE}
\end{figure*}
\begin{algorithm}[t]
\caption{Ring Bayes sparse channel learning}
\label{Sparse Estimation}
\KwIn{$\mathbf{y}_p, \mathbf{\Omega}, \mathbf{S}, \mathbf{A}_R(r,\theta), \tilde{\mathbf{A}}_T(\Phi_T), \epsilon, K_{max}$}
\KwOut{$\hat{\mathbf{H}} = \mathbf{A}_R(r,\theta)\mathrm{vec}^{-1}(\hat{\mathbf{h}}_b) \tilde{\mathbf{A}}_T(\Phi_T)$}

\textbf{Initialization:} $\gamma_i^{(0)} = 1, \forall \: 1 \leq i \leq \lvert \kappa_\theta \rvert \lvert \kappa_r \rvert G_T, \: \hat{\mathbf{\Gamma}}^{(0)} = \mathbf{I}_{\lvert \kappa_\theta \rvert \lvert \kappa_r \rvert G_T}, \: \hat{\mathbf{\Gamma}}^{(-1)} = \mathbf{0}_{\lvert \kappa_\theta \rvert \lvert \kappa_r \rvert G_T}, \: j=0$

\While{$(\parallel\hat{\mathbf{\Gamma}}^{(j)}-\hat{\mathbf{\Gamma}}^{(j-1)} \parallel_\mathcal{F} > \epsilon$ \textrm{and} $j < K_{max})$}{

$j \leftarrow j+1$

\textbf{E-step:} Calculate the \textit{a posteriori} mean and covariance
\begin{equation}
    \begin{aligned}
    \mathbf{\Sigma}^{(j)} &= \big[\mathbf{\Omega}^H\mathbf{S}^{-1}\mathbf{\Omega}+(\hat{\mathbf{\Gamma}}^{(j-1)})^{-1}\big]^{-1}\\
    \boldsymbol{\mu}^{(j)} &= \mathbf{\Sigma}^{(j)}\mathbf{\Omega}^H\mathbf{S}^{-1}\mathbf{y}_p \notag
\end{aligned}
\end{equation}

\textbf{M-step:} Update the hyperparameters 

\For{$i = 1, \cdots, \lvert \kappa_\theta \rvert \lvert \kappa_r \rvert G_T$}{
$\hat{\gamma}_i^{(j)} = \mathbf{\Sigma}^{(j)}(i,i) + |\boldsymbol{\mu}^{(j)}(i)|^2$}
}
\textbf{return:} $\hat{\mathbf{h}}_b = \boldsymbol{\mu}^{(j)}$

\end{algorithm}
Upon convergence, the Bayes estimate of the beamspace channel can be given as
\begin{align}
    \hat{\mathbf{h}}_b = \boldsymbol{\mu}^{(j)}. 
\end{align}
Algorithm-\ref{Sparse Estimation} summarizes the Ring Bayes framework for estimating the sparse mmWave hybrid MIMO channel. The next section discusses the simulation results.
\section{Simulation Results}
To quantify the performance of the proposed Ring Bayes sparse framework, Table-\ref{system_para} outlines the parameters considered.
\begin{table}
\centering
\caption{Simulation parameters for the sparse channel estimate} 
\label{system_para}
\resizebox{0.42\textwidth}{!}{%
\begin{tabular}{|l|c|c|}\hline
\textbf{Parameter} & \textbf{Symbol} & \textbf{Value}  \\\hline
$\#$ of BS antennas & $N_R$ &  $64$  \\\hline
$\#$ of user antennas & $N_T$ & $8$ \\\hline
$\#$ of BS RF chains & $N_{\mathrm{RF}}^R$ &  $16$  \\\hline
$\#$ of user RF chains & $N_{\mathrm{RF}}^T$ &  $2$ \\\hline
$\#$ of frames & $M$ & $20$  \\\hline
$\#$ of scatterers & $N_{cl}$ &  $5$ \\\hline
BS grid size & $\lvert \kappa_\theta \rvert$,$\lvert \kappa_r \rvert$  &  $125$,$4$  \\\hline
User grid size & $G_T$ &  $16$  \\\hline
Operating frequency & $f_c$ &  $30$\:GHz  \\\hline
Concentric-codebook threshold & $\Delta$ & $0.403$   \\\hline
Stopping parameter & $\epsilon, K_{max}$ & $1,30$   \\\hline
\end{tabular}%
}
\end{table}
Note that, as discussed in Section-\ref{an_sample}, we consider $\beta = 2.4048$, which corresponds to $J_0(\beta) \approx 0,$ and $\mathbf{a}^H(r_i,\theta_j) \mathbf{a}(r_i,\theta_j) \leq \Delta \approx 0.403$ for $i \neq j$. Moreover, we compare the performance of our proposed Ring Bayes technique with the existing sparse learning approaches such as OMP \cite{yang2025near}, FOCUSS \cite{bhat2016sparsity}. Fig. \ref{NMSE}(a) illustrates the NMSE performance as a function of SNR. The poor performance of the least squares (LS) \cite{heath2016overview} method is attributed to its inability to incorporate sparsity priors into the estimation process, which poses a significant limitation compared to CS-based approaches. Moreover, the OMP algorithm \cite{yang2025near} suffers from both structural and convergence errors. Specifically, using a low stopping threshold results in a large number of iterations and excessive non-zero components, introducing structural error. Conversely, a high stopping threshold leads to fewer iterations and fails to capture all dominant channel components, resulting in convergence error. On the other hand, the MFOCUSS framework \cite{bhat2016sparsity} is prone to convergence toward suboptimal local minima, which leads to estimation inaccuracies. The proposed Ring Bayes framework leverages the EM algorithm for hyperparameter learning. It eliminates the requirement for manual adjustment of regularization parameters and demonstrates robustness with respect to the dictionary matrix selection. 

Fig. \ref{NMSE}(b) depicts the accuracy of data detection achieved by employing a minimum mean squared error (MMSE) \cite{heath2016overview} receiver with the proposed Ring Bayes framework, along with other conventional sparse estimation based techniques and a Genie-aided detector with perfect CSI as benchmarks. For this, employing the estimated CSI, the precoders and combiners are designed suitably to maximize the beam-focusing gains, the details of which are omitted here due to space constraints. As illustrated, the resulting BER decreases consistently with increasing SNR, and BER corresponding to Ring Bayes estimates approaches that of Genie-aided detector, which closely aligns with the NMSE results presented previously, underscoring the high quality of the CSI estimated by the Bayes technique. Fig. \ref{NMSE}(c) depicts the NMSE versus SNR performance for different numbers of pilot blocks $M$. As $M$ increases, the NMSE consistently improves due to the availability of more measurements, which strengthens the ring Bayes process by accurately capturing the sparse channel structure. This results in reduced estimation uncertainty and improved robustness to noise.
\section{Conclusion}
 We have proposed a novel Ring-Bayes sparse estimation technique for mmWave hybrid MIMO systems employing UCA at BS and ULA at user. Our framework introduces a near-field concentric-ring codebook, in which beamfocusing vectors are strategically sampled at intervals determined by the zeros of Bessel functions to minimize coherence. This sampling strategy effectively captures channel characteristics across both the angular and distance domains, enabling sparse representation of the channel. Simulation results demonstrate significant performance improvements over traditional algorithms, highlighting the efficacy of the proposed approach for advanced mmWave communication systems.
\section{Acknowledgment}
The authors would like to thank the IEEE SPS Scholarship, Telecom Technology Development Fund (TTDF) under Grant TTDF/6G/368 and Anusandhan National Research Foundation’s Grant PM-ECRG/2024/478/ENS.
\bibliographystyle{IEEEtran}
\bibliography{physical_layer_sec.bib}
\end{document}